\documentclass[showpacs,amsmath,amssymb,superscriptaddress]{revtex4}

\usepackage{epsfig}

\usepackage{amssymb}

\begin{document}

\title
{Sigma Exchange in the Nonmesonic Decays of Light Hypernuclei
and Violation of the $\Delta I=1/2$ Rule}

\author{K.~Sasaki} \email{kenjis@post.kek.jp}
\affiliation{Institute of Particle and Nuclear Studies, KEK, 
                 1-1, Oho, Tsukuba, 305-0801 Ibaraki, Japan}
\author{M.~Izaki}
\affiliation{Department of Physics, H27, Tokyo Institute of Technology,
                 Meguro, Tokyo, 152-8551, Japan}
\author{M.~Oka}
\affiliation{Department of Physics, H27, Tokyo Institute of Technology,
                 Meguro, Tokyo, 152-8551, Japan}

\date{\today}

\begin{abstract}
Nonmesonic weak decays of $s$-shell hypernuclei are analyzed in
microscopic models for the $\Lambda N \to NN$ weak interaction.
A scalar-isoscalar meson, $\sigma$, is introduced and its importance
in accounting the decay rates, $n/p$ ratios and proton asymmetry
is demonstrated.
Possible violation of the $\Delta I=1/2$ rule in the nonmesonic weak 
decay
of $\Lambda$ is discussed in a phenomenological analysis and several
useful constraints are presented.  The microscopic calculation shows 
that
the current experimental data indicate a large violation of the $\Delta 
I=1/2$
rule, although no definite conclusion can be derived due to large 
ambiguity
of the decay rate of $^4_{\Lambda}$H.
\end{abstract}


\maketitle

\section{Introduction}

Study of nonmesonic weak decays (NMWD) of $\Lambda$ hypernuclei
 is one of the major subjects of hypernuclear physics.
Dominant contribution in NMWD is known to come from
 $\Lambda N \to NN$ transition in nuclear medium,
 which is a new type of hadronic weak interaction.
It is expected to provide us with valuable information on the weak interaction
 of quarks that may not be available in weak decays of hadrons.
Recent progress in experimental research of NMWD of various hypernuclear
 systems enables us to make quantitative comparison of theoretical predictions
 \cite{CHK:PRC,McK:PRC,Ban:PTP,Ose:NPA,Dub:NPA,ban:ijm,Shm:NPA,ITO:NPA,Mal:PhL,Ram01:PRC,IUM:NPA,Dub:AnP,Par:PRC,IOMI:NPA,SIO01:NPA,Alb:NPA,Jid:NPA,SIO02:NPA}
 and experimental data
 \cite{Szy:PRC,hno:psg,Out:NPA,Zep:NPA,Bha:PRL,aji:prl,Has:PRL,i4}.
During such studies, several interesting discrepancies have been revealed.

One of the puzzling features is the so-called $n/p$ problem, in which 
 the ratio of the $\Lambda n \to nn$ decay rate, $\Gamma_n$ to 
 the $\Lambda p \to pn$ decay rate, $\Gamma_p$, is underestimated
 in the simple-minded one-pion-exchange (OPE) weak interaction.
From theoretical point of view, 
 the essence of this puzzle is attributed to the strong tensor force brought by 
 OPE.
The $n/p$ ratio is strongly suppressed (to about $0.1$) 
 by enhancement of the $\Gamma_p$ due to the tensor force of OPE.
Recent experimental data, however, established that the $n/p$ ratio is around 1/2
 for $^5_{\Lambda}$He and $^{12}_{\Lambda}$C.
In the previous studies
 \cite{ITO:NPA,IOMI:NPA,SIO01:NPA,SIO02:NPA}, 
 we found that the total decay rates and the $n/p$ ratios
 are both sensitive to short-range components of the baryonic weak interaction,
 which are represented by one-kaon-exchange (OKE), and the direct-quark (DQ) transition.
We showed in the OPE+OKE+DQ model that
 the $\Gamma_n$ is enhanced due to the short-range contributions and thus can reproduce
 the observed $n/p$ ratio both in nuclear matter and in light hypernuclei.
At the same time, we found that the total decay rates of light hypernuclei tend
 to be overestimated.

Another quantity that shows discrepancy between experiment and theory
 is asymmetry of emitted proton from polarized hypernuclei.
Recent theoretical predictions \cite{Par:PRC,SIO01:NPA,SIO02:NPA,Ito:Pdis}
 yield large negative values of the asymmetry parameter, $\alpha$, 
 while new experimental data suggest a smaller positive asymmetry for $^5_{\Lambda}$He decay
 \cite{i4}.
The asymmetry comes from interference between the parity conserving (PC) part and 
 the parity violating (PV) part of the decay amplitudes, and thus is sensitive to the detail
 decomposition of the decay amplitudes.
In other words, it has more discriminative power to
 determine goodness of the models than the decay rates.

Besides the calculations with microscopic models,
 an analysis employing an effective field theory (EFT) was carried out recently \cite{Jun98,Par:hep}.
There the short range parts of the interactions are represented by four-point baryonic operators.
By fitting strength parameters to current experimental data, including the proton asymmetry,
 it was shown that the largest term comes from the isospin- and spin-independent central operator.
Thus the EFT approach suggests that in order to reproduce the proton asymmetry data, 
 the microscopic models should be supplemented by central interactions.

Hinted by this observation, we consider scalar-meson exchange in
 the weak $\Lambda N \to NN$ transition. 
The scalar $\sigma$ meson with $I=0$ has been introduced in the context of chiral symmetry 
 of QCD.  
When the symmetry is spontaneously broken due to non-zero quark condensate,
 the pion, $\pi$, appears as a (pseudo) Nambu-Goldstone boson, while its chiral partner
 is a scalar-isoscalar $\sigma$ for the $N_f=2$ chiral symmetry.
Although the picture of chiral symmetry breaking of QCD has been established for some time,
 existence of $\sigma$ as a real meson was not confirmed until recently.  
It appears as a broad resonance in the $\pi-\pi$ scattering phase shift,
 and its mass happens to be around 600 MeV \cite{PDG:PLB}.
It has been long known that strong $NN$ potential requires the $\sigma$ exchange
 in order to obtain enough attraction both in the spin singlet and triplet channels.
As the $\sigma$ mass is of the same range as the kaon,
 it is natural to expect its significant role in the weak baryonic interaction as well.  
Thus we consider the one-sigma exchange (OSE)
 in $\Lambda N\to NN$ transition.

Our model, which we call DQ+, now consists of OPE, OKE, OSE and DQ.  
We will show, in this paper, that the model contains necessary features
 in order to reproduce all the experimental data
 on NMWD of light hypernuclei, and indeed the proton asymmetry puzzle can be solved by
 the contribution of OSE.

Another interesting property of the strangeness changing weak interactions of hadrons
 is its isospin property.  
It is well known that the decays of kaon and hyperons satisfy 
 so-called $\Delta I=1/2$ rule, which indicates the dominance of the $I=1/2$ transition operator
 to the $I=3/2$ operator.  
In the standard theory of the weak interaction of quarks,
 the transition $s+\bar u \to W^- \to d+\bar u$ allows both $\Delta I=1/2$ and 3/2. 
Yet, in the hadron decays, the $\Delta I=3/2$ transition is much weaker than the $\Delta I=1/2$.
The ratio of the amplitude is typically $20$ in the decays of $K$ and hyperons.
The origin of this empirical ``rule'' is not completely understood.
In $K\to \pi\pi$ decays, $\Delta I=1/2$ dominance may be explained by contribution of
 the scalar-isoscalar meson, $\sigma$, in the $s$-channel \cite{Ino03:PTP}.
The enhancement comes due to the closeness of the masses of $K$ and $\sigma$.
Suppression of $\Delta I=3/2$ transition in the baryon weak decays may be explained by
color structure of quark model wave function of the baryon (MMPW theorem)
 \cite{Miu:PTP,Pat:PRD}.
These mechanisms are rather specific to the particular decays and are not generalized to
 the nonmesonic weak decays, $YN\to NN$.

It is therefore important and interesting to test whether the $\Delta I=1/2$ rule 
 is also effective in NMWD of hypernuclei.  
This is the second purpose of this paper.
We note that the meson exchange processes are all dominated by the $\Delta I=1/2$ amplitudes.  
First OPE is assumed to be purely $\Delta I=1/2$ because the $\pi\Lambda N$
 weak vertex causes free $\Lambda$ decay. 
From the $\Delta I=1/2$ dominance of the $\Lambda\to N\pi$ decays, we expect that
 the vertex is (almost) purely $\Delta I=1/2$.
OKE is also supposed to have only $\Delta I=1/2$, because the $KNN$ weak coupling is
 derived from the $\pi \Lambda N$ coupling using the SU(3) relation.
It is obvious that OSE, or the weak $\sigma \Lambda N$ coupling, is also purely 
 $\Delta I=1/2$, because the isospin of $\sigma$ is zero.

In contrast, the direct quark (DQ) process may contain $\Delta I=3/2$ transitions.
We employ the effective weak lagrangian derived from the standard theory
 with one-loop QCD corrections \cite{Pas:NPB}.  
The preturbative QCD corrections, which are valid only at the momentum scale of $M_W$,
 are ``improved'' by using the QCD renormalization group equation.  
The resulting effective lagrangian is given in terms of four-quark local operators, 
 such as $(\bar d_L u_L) (\bar u_L s_L)$.
A part of the $\Delta I=1/2$ enhancement (and $\Delta I=3/2$ suppression)
 is included in the course of the down-scaling according to the renormalization group 
 equation, but certain $\Delta I=3/2$ strength still remains
 \cite{Alt:PhL,Vai:JETP}.
The DQ transition potential thus contains $\Delta I=3/2$ part. 
In the previous study, we predicted significant violation of the $\Delta I=1/2$ rule 
 in the $J=0$ transition amplitudes in particular.

In this paper, we consider how the $\Delta I=3/2$ transition affects the transition 
 rates of light hypernuclei and check the validity of $\Delta I=1/2$ rule 
 within the available experimental data.

This paper is organized as follows.
In sect.~2, we summarize the formulation of the weak transition calculations.
In sect.~3, several general relations based on simple parameterization of the
 decay rates of the $s$-shell hypernuclei are given and the validity of $\Delta I=1/2$
 rule is considered.
In sect.~4, we introduce the $\sigma$ exchange and complete our DQ+ model.
The weak coupling parameters for the $\sigma$ meson are determined so as to
 reproduce the data from the $s$-shell hypernuclei.  
We give the full results including the proton asymmetry parameter and point out
 the important roles of the $\sigma$ meson exchanges.
Conclusions are given in sect.~5.

\section{Decay rates of light hypernuclei}
Observables of the weak decay of $s$-shell hypernuclei give us 
 a chance to discuss the properties of the $\Lambda N \to NN$ interaction,
 the $\Gamma_{n}/\Gamma_{p}$ ratio and the $\Delta I = 1/2$ dominance.
Block and Dalitz~\cite{BD:PRL} performed
 an analysis based on the lifetime data of light hypernuclei,
 which were updated by some other authors~\cite{Dov:FSS,Sch:NPA}.
For $s$-shell hypernuclei,
 the initial $\Lambda N$ system can be assumed to be in the relative
 {\it{s}}-wave state, and we consider the $\Lambda N \to NN$ transition with
 the six $^{2S+1}L_J$ combinations listed in Table~\ref{TAB:amp}.

\begin{table}[tb]
 \begin{center}
 \caption{Possible $^{2S+1}L_J$ combinations and amplitudes
          for nonmesonic weak transitions of the $s$-shell hypernuclei.}
\begin{tabular}{cccccc}
\hline \hline
\multicolumn{2}{c}{State} & \multicolumn{1}{c}{Parity} & Isospin & \multicolumn{2}{c}{amplitude} \\
\cline{1-2}\cline{5-6}
Initial & Final &  &  & $I^f_z=0$ & $I^f_z=-1$ \\
\hline
${^1S_0}$ & ${^1S_0}$ & PC & $I^f =1$ & $a_p$ & $a_n$ \\
          & ${^3P_0}$ & PV & $I^f =1$ & $b_p$ & $b_n$ \\
${^3S_1}$ & ${^3S_1}$ & PC & $I^f =0$ & $c_p$ & ---   \\
          & ${^3D_1}$ & PC & $I^f =0$ & $d_p$ & ---   \\
          & ${^1P_1}$ & PV & $I^f =0$ & $e_p$ & ---   \\
          & ${^3P_1}$ & PV & $I^f =1$ & $f_p$ & $f_n$ \\
\hline \hline
\end{tabular}
 \end{center}
\label{TAB:amp}
\end{table}

Using amplitudes $a_p \sim f_n$,
 we can express the total decay rates of light hypernuclei
 in a short-handed notation:
\begin{eqnarray}
&&
 \Gamma_{NM}({^5_\Lambda}{\rm{He}}) = 
   |a_p^5|^2 + |b_p^5|^2
 + 3\left( |c_p^5|^2 + |d_p^5|^2 + |e_p^5|^2 +|f_p^5|^2 \right)
\nonumber \\
&& \hspace*{21.5mm}
 + |a_n^5|^2 + |b_n^5|^2
 + 3|f_p^5|^2
\\
&&
 \Gamma_{NM}({^4_\Lambda}{\rm{He}}) = 
   |a_p^4|^2 + |b_p^4|^2 + 3\left( |c_p^4|^2 + |d_p^4|^2 + |e_p^4|^2 +|f_p^4|^2 \right)
\nonumber \\
&& \hspace*{21.5mm}
 + 2\left( |a_n^4|^2 + |b_n^4|^2 \right)
\\
&&
 \Gamma_{NM}({^4_\Lambda}{\rm{H}}) \phantom{e} = 
   2\left( |a_p^4|^2 + |b_p^4|^2 \right)
 + |a_n^4|^2 + |b_n^4|^2  + 3|f_p^4|^2
\end{eqnarray}
where the superscript indicates the mass number of hypernucleus.
In a same way,
 the $n/p$ ratios of light hypernuclei are written down:
\begin{eqnarray}
&&
 \frac{\Gamma_n}{\Gamma_p}({^5_\Lambda}{\rm{He}}) =
  \frac{|a_n^5|^2 + |b_n^5|^2 + 3|f_p^5|^2}
   {|a_p^5|^2 + |b_p^5|^2 + 3\left( |c_p^5|^2 + |d_p^5|^2 + |e_p^5|^2 +|f_p^5|^2 \right)}
\\
&&
 \frac{\Gamma_n}{\Gamma_p}({^4_\Lambda}{\rm{He}}) =
  \frac{2\left( |a_n^4|^2 + |b_n^4|^2 \right)}
   {|a_p^4|^2 + |b_p^4|^2 + 3\left( |c_p^4|^2 + |d_p^4|^2 + |e_p^4|^2 +|f_p^4|^2 \right)}
\\
&&
 \frac{\Gamma_n}{\Gamma_p}({^4_\Lambda}{\rm{H}}) \phantom{e} =
   \frac{|a_n^4|^2 + |b_n^4|^2  + 3|f_n^4|^2}{2\left( |a_p^4|^2 + |b_p^4|^2 \right)}.
\end{eqnarray}
The asymmetry parameter~\cite{Nab:PRC} is obtained  by 
\begin{eqnarray}
  \alpha = \frac{ 2 (- \sqrt{3}[a_p^5 e_p^5] - [b_p^5 c_p^5] + \sqrt{2}[b_p^5 d_p^5] 
    + \sqrt{6} [c_p^5 f_p^5] + \sqrt{3}[d_p^5 f_p^5] ) }
      { |a_p^5|^2 + |b_p^5|^2 + 3\left( |c_p^5|^2 + |d_p^5|^2 + |e_p^5|^2 +|f_p^5|^2 \right) }
\label{EQ:asy}
\end{eqnarray}
 where we define $[a_p e_p] \equiv {\rm{Re}} (a^\ast_p e_p)$, etc.
Note that there appear interference terms between 
 the $J=0$ and $J=1$ amplitudes, 
 such as $[a_p e_p]$ and $[b_p c_p]$, in eq.(\ref{EQ:asy}).

The $\Delta I =1/2$ rule for the $\Lambda N \to NN$ transition leads to 
 the isospin relations:
\begin{eqnarray}
     a_n  =  \sqrt{2} a_p, \hspace{2mm}
     b_n  =  \sqrt{2} b_p, \hspace{2mm} {\rm{and}} \hspace*{2mm}
     f_n  =  \sqrt{2} f_p 
\label{EQ:DelI12}
\end{eqnarray}
 for the decay amplitudes listed in Table \ref{TAB:amp}.
This rule also makes
\begin{eqnarray}
 \kappa &\equiv&
 \frac{\Gamma_n ({^4_\Lambda}{\rm{He}})}{\Gamma_p ({^4_\Lambda}{\rm{H}})}
 = \frac{|a_n^4|^2 + |b_n^4|^2}{|a_p^4|^2 + |b_p^4|^2}
 ,
\end{eqnarray}
to be equal to $2$.
Therefore the $\kappa$ is important to check the validity of the $\Delta I=1/2$ rule
 for the $\Lambda N \to NN$ transition from both the theoretical and experimental
 points of view.

In the present analysis, we did not include effects of virtual $\Sigma$ mixing
 in hypernuclei.
Importance of the $\Sigma$ mixing in ${^4_\Lambda}$He and ${^4_\Lambda}$H 
was pointed out elsewhere
 \cite{SIO02:NPA,GGW:PRC,AHSM:PRL,HKMYY:NPA,NKG:PRL,{NAS:PRL}}.
Even if the microscopic interactions preserve the $\Delta I=1/2$ rule,
the $\Lambda-\Sigma$ mixing may result in deviation from the $\Delta I=1/2$ relation:
$ \kappa = 2 $.
We also neglect decay amplitudes which are induced by two nucleons, i.e.,
$\Lambda NN\to NNN$ decays. This is another process which may modify the 
above relations.
Thus, strictly speaking, we need to subtract those extra contributions before 
applying the above relations to the experimental data.

The $\Lambda N \to N N$ transition rate is given by
\begin{equation}
\Gamma_{N} = 
 \int \! \! \frac{d^3{p_1'}}{(2 \pi)^3} 
 \int \! \! \frac{d^3{p_2'}}{(2 \pi)^3}
 \frac{1}{2J_H+1} 
 \sum_{i,f} (2 \pi) \delta(E_f - E_i)
 | M_{fi} |^2
\label{EQ:decayrNR}
\end{equation}
where $M_{fi}$ is the $\Lambda N \to N N$ transition amplitude, 
 $J_H$ is the total spin of initial hypernucleus, and 
 ${p'}_1$ and ${p'}_2$ are momenta of emitted particles, {\it{i.e.}}, hyperon and nucleon.
The summation indicates a sum over all quantum numbers of the initial
 and final particle systems.

After the decomposition of angular momentum, the explicit form of $|M_{fi}|^2$ is 
\begin{equation}
|M_{fi}|^2 =
 (4 \pi)^4 \left|
 \int \! \! \! \! \int \! \! \! \! \int
 \Psi^{L'S'J}_f(R, r')
 {V}^{L L'}_{S S'J}(r,r')
 \Psi^{LSJ}_i(R, r)
 r^2dr {{r'}^2}dr' R^2dR
 \right|^2
\label{EQ:nonrelM}
\end{equation}
where ${V}^{L L'}_{S S'J}(r,r')$
 is the (non-local) transition potential and
 $\Psi^{LSJ}(R, r)$ 
 is the wave function of the $\Lambda N$ or $NN$ two-body system
 in the configuration space.
The indices $L$, $S$, and $J$ indicate
 the orbital angular momentum, spin, and total spin for two-body system, respectively.

We take the wave function of the $\Lambda$-$N$ two body systems in the form,
\begin{eqnarray}
  \phi_Y({\vec{r}}_Y) \phi_N({\vec{r}}_N)
      \left[
        \left(
        1 - e^{-r^2 / a^2}
        \right)^n
      - br^2 e^{-r^2 / c^2}
      \right]
\end{eqnarray}
 where $\phi_i$ stands for the single-particle wave function inside the nucleus,
 and $r=|\vec r_Y - \vec r_N |$.
For $\phi_N$,
 we assume the harmonic oscillator shell model,
 and the size parameter is chosen so as to reproduce
 the size of the nucleus without $\Lambda$.
The $\phi_\Lambda$ is described by the solution of the Schr\"odinger equation 
 with a $\Lambda$-core potential~\cite{IOMI:NPA}.
The parameters for the short-range correlation
 are $a= 0.5$, $b=0.25$, $c=1.28$ and $n=2$,
 which reproduce the realistic $\Lambda$-$N$ correlation~\cite{Par:PRC}.

The wave function of the final two nucleons emitted in the two-body 
 weak process is assumed to be the plane wave with short-range correlation:
\begin{eqnarray}
 e^{i \vec K \cdot \vec R'} e^{i \vec k \cdot \vec r'}
        \left[
        1 - j_0(q_c r')
        \right]
\end{eqnarray}
 where $\vec r'=\vec r_{N_2} - \vec r_{N_1}$,
 $\vec R'= (\vec r_{N_2} + \vec r_{N_1})/2$
 and $q_c$ = 3.93 [fm$^{-1}$].
This approximation may be justified for light nuclei as the 
 momenta of the emitted nucleons are relatively high ($\sim 400$ MeV/c).

\begin{table}
\caption{The strong and weak coupling constants in the present model.
         The strong couplings are taken from the Nijmegen soft-core potential (NSC97)
         \cite{NSC97:PRC}.
         The weak couplings are given in units of 
         $g_w \equiv G_F m_\pi^2 = 2.21 \times 10^{-7}$.
		 The weak coupling constants for $\sigma$ meson are the values used in sect.~4.}
\begin{center}
\begin{tabular}{|c||llr|llr|llr|}
\hline
\hline
Meson & \multicolumn{3}{c|}{Strong c.c.} & \multicolumn{6}{c|}{Weak c.c.} \\ 
\cline{5-10}
(mass) & & & & \multicolumn{3}{c|}{PC} & \multicolumn{3}{c|}{PV} \\ 
\hline
$\pi$ 
 & $g_{N \! N\pi}$       & $\! \! \! = \! \! \!$ & $13.16$  
 & $B_{\pi^0 \Lambda n}$ & $\! \! \! = \! \! \!$ & $7.15$  
 & $A_{\pi^0 \Lambda n}$ & $\! \! \! = \! \! \!$ & $\! -1.05$ \\
 $(138 {\rm{MeV}})$
 & & & 
 & $B_{\pi^- \Lambda p}$ & $\! \! \! = \! \! \!$ & $\! -10.11$
 & $A_{\pi^- \Lambda p}$ & $\! \! \! = \! \! \!$ & $1.48$ \\
\hline
\raisebox{-1.8ex}[0pt][0pt]{ $K$ }  
 & $g_{\Lambda NK}$      & $\! \! \! = \! \! \!$ & $\! -17.65$ 
 & $B_{K^0 n n}$         & $\! \! \! = \! \! \!$ & $\! -16.19$ 
 & $A_{K^0 n n}$         & $\! \! \! = \! \! \!$ & $2.83$ \\
\raisebox{-1.8ex}[0pt][0pt]{ $(495 {\rm{MeV}})$ }
 & & & 
 & $B_{K^0 p p}$         & $\! \! \! = \! \! \!$ & $6.65$ 
 & $A_{K^0 p p}$         & $\! \! \! = \! \! \!$ & $2.09$ \\
 & & & 
 & $B_{K^+ p n}$         & $\! \! \! = \! \! \!$ & $\! -22.84$ 
 & $A_{K^+ p n}$         & $\! \! \! = \! \! \!$ & $0.76$ \\
\hline
$\sigma$
& $g_{N\!N\sigma}$    & $\! \! \! = \! \! \!$ & $13.16$
 & $A_{\sigma}^{ME}$  & $\! \! \! = \! \! \!$ & $3.8\phantom{0}$  
 & $B_{\sigma}^{ME}$  & $\! \! \! = \! \! \!$ & $1.2\phantom{0}$ \\
 $(550 {\rm{MeV}})$
 & & &
 & $A_{\sigma}^{DQ+}$    & $\! \! \! = \! \! \!$ & $3.9\phantom{0}$  
 & $B_{\sigma}^{DQ+}$    & $\! \! \! = \! \! \!$ & $6.6\phantom{0}$ \\
\hline
\hline
\end{tabular}
\end{center}
\label{TAB:SWcc}
\end{table}

The general form of one-pion exchange (OPE) potential
 for the $\Lambda N \to NN$ transition can be written as
\begin{equation}
 V_{\Lambda N \to NN}(\vec q)
  =
  g_s [\bar{u}_N \gamma_5 u_N]
    {\frac{1}{{\vec q}^2 + {\tilde m}_{i}^2}}
    \left( {  \frac{ \Lambda_{i}^2 -{\tilde m}_{i}^2 }
              { \Lambda_{i}^2 + {\vec q}^2   }
     } \right)^2
  g_w
    [\bar{u}_N (A+B \gamma_5) u_\Lambda]
  \label{eq:OPE}
\end{equation}
 where the coupling constants $g_s$, $g_w$, $A$ and $B$, 
 shown in Table~\ref{TAB:SWcc},
 are chosen properly for each transition.
It is easy to confirm that the weak coupling constants satisfy the $\Delta I=1/2$ conditions,
namely, 
$A_{\pi^- \Lambda p} = -\sqrt{2} A_{\pi^0 \Lambda n}$ and 
$B_{\pi^- \Lambda p} = -\sqrt{2} B_{\pi^0 \Lambda n}$.
A monopole form factor with cutoff parameter, 
 $\Lambda_{\pi}$ = 800 MeV, 
 is employed for each vertex.
As the energy transfer is significantly large, 
 we introduce the effective meson mass:
\begin{equation}
 \tilde{m} = \sqrt{m^2 - (q^0)^2}, 
 \hspace*{5mm}
 q^0 = 
 88.5 {\rm{MeV}}.
\end{equation}

The kaon exchange (OKE) potential can be constructed similarly.
Both the strong and weak coupling constants are evaluated employing the
 assumption of the flavor SU(3) symmetry and they are also listed in Table~\ref{TAB:SWcc}.
The cutoff parameter, $\Lambda_{K} =1300$ MeV, is used for the form factor.
The $\Delta I=1/2$ rule for the weak $KNN$ vertex requires the conditions,
\begin{eqnarray}
A_{K^0 n n} &=& A_{K^0 p p} +A_{K^+ p n} \nonumber \\
B_{K^0 n n} &=& B_{K^0 p p} +B_{K^+ p n},
\end{eqnarray}
which are easily seen to be satisfied.

The third meson considered here is the $\sigma$ meson, which is a scalar and isoscalar
 meson with the couplings,
\begin{eqnarray}
&&
\mathcal{H}_{s}^{\sigma NN} = g_{s} \bar{\psi}_{N}(x)  {\phi}_{\sigma}(x)  \psi_{N}(x)
\nonumber \\
&&
\mathcal{H}_{w}^{\sigma \Lambda N} = g_{w}  \bar{\psi}_{n}(x) 
           (A_{\sigma} + B_{\sigma} \gamma_{5}) \phi_{\sigma}(x) \psi_{\Lambda}(x) .
\end{eqnarray}
The weak hamiltonian $\mathcal{H}_{w}$ consists of
 a parity conserving part (proportional to $A_{\sigma}$)
 and a parity violating part ($B_{\sigma}$). 
We employ $550$ MeV for the mass of $\sigma$ and $1200$ MeV for the cutoff mass.
From the medium-range attraction in the nuclear force potential, 
 the strong coupling constant is known to be about 10 but, here, is taken to be
 same as the strong $\pi NN$ coupling strength, i.e., $g_{\sigma NN} = g_{\pi NN}$.
The results do not depend on the choice of $g_{\sigma NN}$, because it is always
multiplied by the weak coupling constants, $A_{\sigma}$ or $B_{\sigma}$, which 
are free parameters in the present analysis.
As was mentioned already, the $\sigma \Lambda N$ coupling contains only
 $\Delta I=1/2$ transition because the $\sigma$ meson is isoscalar.
Unlike the OPE and OKE,
 this potential does not include the tensor transition potential in 
 a parity conserving channel.
Hence
 the OSE cannot affect the ${^3S_1}\to{^3D_1}$ channel ($d_p$).

The DQ potential is given as a nonlocal form as
\begin{equation}
   {V_{DQ}}^{L L'}_{S S'J}(r,r')
    =  -{G_{F}\over\sqrt{2}} \times W \, \sum_{i=1}^{7}
    \left\{ V_{i}^f f_{i}(r, r') + V_{i}^g g_{i}(r, r') 
    + V_{i}^h h_{i}(r, r') 
    \right\}
\end{equation}
where $r$ ($r'$) stands for the radial part of the relative coordinate 
 in the initial (final) state.
The explicit forms of $f_i$, $g_i$, and $h_i$ are given in ref. \cite{SIO01:NPA},
 and the coefficients, $V^k_{i}$, for the $\Lambda N \to NN$ transitions are also
 listed in ref. \cite{SIO01:NPA}.

%
\section{Phenomenological Analysis}
Recently, Alberico {et al} \cite{AGa:PLe} carried out an analysis 
 of experimental data of the NM decays of the $s$-shell hypernuclei
 from the viewpoint of validity of the $\Delta I = \frac{1}{2}$ rule. 
Comparing analyses with and without the constraint from $\Delta I = \frac{1}{2}$,
 they found that the current experimental data cannot confirm nor deny its validity.
We here follow their analysis with new experimental data and study 
 qualitative features of the decay rates in specific flavor and spin channels.

The data employed in the analyses are summarized in Table~\ref{ta1},
 where we use the new data of the total NM decay rate and
 $\gamma\equiv\Gamma_{n}/\Gamma_{p}$ ratio of ${^5_\Lambda}{\rm{He}}$
 taken from Ref.~\cite{i4}, and compare the results  with those obtained from the old data
 in Ref.~\cite{Szy:PRC}.
Among these data,
 the nonmesonic decay rate of ${^4_\Lambda}{\rm{H}}$ is the most ambiguous one.
We here take the weighted average of a recent estimate by Outa \cite{Out:NPA}
 and an old estimate by Block and Dalitz \cite{BD:PRL},
 but it should be noted that
 these numbers were not obtained by direct measurements,
 but were estimated with theoretical assumptions.

\begin{table}
\caption{Experimental data employed in the analysis in sec.~3, in units of $\Gamma^{free}_{\Lambda}$.
The Set I is the data employed in an analysis by Ref~\cite{AGa:PLe}, and the Set II is those for the present analysis.}
\label{ta1}
\begin{center}
\begin{tabular}{|l|c|c|}
\hline
\hline
  & Set I    & Set II  \\
\hline
   $\Gamma_{NM} ({^4_\Lambda}{\rm{H}})$ 
  & $0.22 \pm 0.09$ \cite{AGa:PLe} & $0.22 \pm 0.09$ \cite{AGa:PLe}  \\
\hline
   $\Gamma_{NM} ({^4_\Lambda}{\rm{He}})$  
  & $0.20 \pm 0.03$ \cite{Zep:NPA} & $ 0.20 \pm 0.03$ \cite{Zep:NPA} \\
\hline
   $\gamma_4^{\rm{He}} = \frac{\Gamma_{n}}{\Gamma_{p}} ({^4_\Lambda}{\rm{He}})$ 
  & $0.25 \pm 0.13$ \cite{Zep:NPA} & $0.25 \pm 0.13$ \cite{Zep:NPA}  \\
\hline
   $\Gamma_{NM} ({^5_\Lambda}{\rm{He}})$   
  & $0.41 \pm 0.14$ \cite{Szy:PRC} & $0.395 \pm 0.016$  \cite{i4} \\
\hline
   $\gamma_5 =\frac{\Gamma_{n}}{\Gamma_{p}} ({^5_\Lambda}{\rm{He}})$ 
  & $0.93 \pm 0.55$ \cite{Szy:PRC} & $0.44 \pm 0.11$  \cite{i4}  \\
\hline 
\hline
\end{tabular}
\end{center}
\end{table}

We assume that the nonmesonic decay rates of the $s$-shell hypernuclei are 
 parameterized as
\begin{eqnarray}
&&
\Gamma_{NM}({^4_\Lambda}{\rm{H}})\phantom{e} 
= \Gamma_{p}({^4_\Lambda}{\rm{H}})+ \Gamma_{n}({^4_\Lambda}{\rm{H}}) \label{eqa4}\\
&& 
\quad\Gamma_{p}({^4_\Lambda}{\rm{H}})=
\frac{\bar\rho_{4}}{6} 2R_{p0},
\qquad\Gamma_{n}({^4_\Lambda}{\rm{H}})=
\frac{\bar\rho_{4}}{6} (R_{n0} + 3R_{n1} )   \nonumber \\
&&
\Gamma_{NM}({^4_\Lambda}{\rm{He}}) 
= \Gamma_{p}({^4_\Lambda}{\rm{He}})+ \Gamma_{n}({^4_\Lambda}{\rm{He}}) \label{eqa4e}\\
&&
\quad\Gamma_{p}({^4_\Lambda}{\rm{He}})=
\frac{\bar\rho_{4}}{6} (R_{p0} + 3R_{p1}), 
\qquad\Gamma_{n}({^4_\Lambda}{\rm{He}})=
\frac{\bar\rho_{4}}{6} 2R_{n0} \nonumber \\
&&
\Gamma_{NM}({^5_\Lambda}{\rm{He}}) 
= \Gamma_{p}({^5_\Lambda}{\rm{He}})+ \Gamma_{n}({^5_\Lambda}{\rm{He}}) \label{eqa5}\\
&&
\quad\Gamma_{p}({^5_\Lambda}{\rm{He}})=
\frac{\bar\rho_{5}}{8} (R_{p0} + 3R_{p1}),
\qquad\Gamma_{n}({^5_\Lambda}{\rm{He}})=
\frac{\bar\rho_{5}}{8} (R_{n0} + 3R_{n1}),   \nonumber 
\end{eqnarray}
where $R_{NJ}$ are the strengths of the $\Lambda N \to NN$ elementary interactions
 for the spin-singlet ($R_{n0}$ , $R_{p0}$)
 and spin-triplet ($R_{n1}$ , $R_{p1}$) channels.
They are related to the $a\sim f$ amplitudes by
\begin{eqnarray}
&&
   |a_p^A|^2 + |b_p^A|^2 = \frac{\bar\rho_{A}}{2(A-1)} R_{p0}, 
\nonumber \\
&&
   |a_n^A|^2 + |b_n^A|^2 = \frac{\bar\rho_{A}}{2(A-1)} R_{n0},
\nonumber \\
&&
   |c_p^A|^2 + |d_p^A|^2 + |e_p^A|^2 +|f_p^A|^2
                         = \frac{\bar\rho_{A}}{2(A-1)} R_{p1},
\nonumber \\
&&
   |f_n^A|^2             = \frac{\bar\rho_{A}}{2(A-1)} R_{n1}.
\label{eq4}
\end{eqnarray}
The coefficient $\bar\rho_{A}$ denotes the average nucleon density 
 at the position of $\Lambda$ defined by 
\begin{eqnarray}
\bar\rho_A \equiv \int d\vec r\, \rho_A(\vec r) |\psi_{\Lambda}(\vec r)|^2  .
\end{eqnarray}

There is an interesting theorem derived from the parameterization, Eq.~(\ref{eq4}).
Define
\begin{eqnarray}
 R_4 &\equiv&
 \frac{\Gamma_{NM} ({^4_\Lambda}{\rm{H}})}{\Gamma_{NM} ({^4_\Lambda}{\rm{He}})}
 \nonumber
\end{eqnarray}
and then it is straightforward to prove the following theorem,
 using the fact that all the $R_{NJ}$'s are positive.

\medskip
{\bf Theorem:}
\begin{eqnarray}
{\rm Min} (\gamma_5, \kappa^{-1})  < R_4 < {\rm Max} (\gamma_5, \kappa^{-1})
\end{eqnarray}

\smallskip

\noindent 
This theorem is extremely important because the ratio $\kappa$ is
 directly related to $\Delta I$ in the weak transition.
Namely, $\kappa$ is determined solely by the Clebsch-Gordan coefficient 
 when the isospin of the transition operator, $\Delta I$, is purely 1/2 or 3/2:
\begin{eqnarray}
\kappa &=& \frac{\Gamma_n ({^4_\Lambda}{\rm{He}})}{\Gamma_p ({^4_\Lambda}{\rm{H}})}
        = \frac{R_{n0}}{R_{p0}} = 
 \left\{
\begin{array}{ll}
 2   & \hbox{for $\Delta I=1/2$} \\
 1/2 & \hbox{for $\Delta I=3/2$} 
\end{array}
 \right.
\label{DI1/2-0}
\end{eqnarray}

The new data\cite{i4} suggests $\gamma_5 \sim 0.5$.
If we assume $\Delta I=1/2$ or equivalently $1/\kappa = 1/2$, then 
the Theorem restricts $R_4$ to be around 0.5.
The current estimate of $\Gamma_{NM}({^4_\Lambda}{\rm{H}})$ does 
not seem to support $R_4=0.5$, although it is not completely rejected.
In contrast, if we remove the $\Delta I=1/2$ constraint, the Theorem allows 
the two decay rates in $A=4$ to be comparable, i.e., $R_4 \sim 1$, as the central values of
the current estimate indicate.

Now we determine $R_{NJ}$ from the two sets of the experimental data given in Table \ref{ta1}.
We first fix $\bar\rho_5$, again following Ref.~\cite{AGa:PLe} which uses an estimate from a model
wave function, $\bar\rho_{5}=0.045 {\rm{fm}}^{-3}$.
We also use this value throughout the phenomenological analysis in this section.
In fact, the results are not sensitive to the choice of this value.
This leaves five unknown parameters, $\bar\rho_4$ and four $\Gamma_{NJ}$, which
can be determined by the five experimental data tabulated in Table \ref{ta1}.
In particular, the density parameter $\bar\rho_4$ can be determined by the relation,
\begin{eqnarray}
\frac{\Gamma_{p}({^5_\Lambda}{\rm{He}})}{\Gamma_{p}({^4_\Lambda}{\rm{He}}) } =
  \frac{3 \bar\rho_{5}}{4 \bar\rho_{4}},
\end{eqnarray}
where $\Gamma_{p}(_{\Lambda}Z)$ is obtained by
\begin{eqnarray}
\Gamma_{p}(_{\Lambda}Z) =
   \Gamma_{NM} (_{\Lambda}Z) (1+\gamma(_{\Lambda}Z))^{-1} . 
\end{eqnarray}
We obtain $\bar\rho_{4}=$0.026 ${\rm{fm}}^{-3}$  for the data Set I,
while it is 0.020 ${\rm{fm}}^{-3}$ for the Set II.

The decay rates, $R_{NJ}$, determined by the two sets of data in Table~\ref{ta1},
are given in the last two columns of Table~\ref{ta2}.
One sees that for both the data sets the central value of $\kappa$ is much smaller than 2, 
and thus indicates strong violation of the $\Delta I = \frac{1}{2}$ rule.
It is seen that the new data set reduces the error bar very much and makes the conclusion 
prominent.

\begin{table}
\caption{The results of analyses based on Eq.~(\protect\ref{eqa4}, \ref{eqa4e}, \ref{eqa5}) with and without the
$\Delta I=1/2$ constraint.
$\Gamma_{NM}$ are given in units of $\Gamma_{\Lambda}^{free}$ and 
$R_{NJ}$ are in units of ${\rm{fm}}^{3}$.}
\label{ta2}
\begin{center}
\begin{tabular}{|c|c|c|c|c|}
\hline
\hline
\multicolumn{1}{|c|}{}     &   \multicolumn{2}{c|}{with $\Delta I = \frac{1}{2}$ rule} 
         &   \multicolumn{2}{|c|}{without $\Delta I = \frac{1}{2}$ rule}    \\
\cline{2-5}     
              & Set I  & Set II   & Set I   & Set II  \\
\hline         
    $ R_{n0}$ & $4.7 \pm 2.1$  & $6.1 \pm 2.7$  & $4.7 \pm 2.1$        & $6.1 \pm 2.7$   \\
\hline         
    $ R_{p0}$ & $2.3 \pm 1.0$  & $3.0 \pm 1.3$  & $7.9_{-7.9}^{+16.6}$ & $22.8 \pm 14.5$ \\
\hline         
    $ R_{n1}$ & $10.3 \pm 8.6$ & $5.1 \pm 3.0$  & $10.3 \pm 8.6$       & $5.1 \pm 3.0 $  \\
\hline         
    $ R_{p1}$ & $11.5 \pm 6.7$ & $15.2 \pm 3.1$ & $9.8 \pm 5.5$        & $8.7 \pm 4.8$   \\
\hline         
\hline
    $\kappa= R_{n0}/R_{p0}$
              & 2              &  2             & $0.6 _{-0.6}^{+1.3}$ & $0.27 \pm 0.21$ \\            
\hline
\hline
   $\Gamma_{NM} ({^4_\Lambda}{\rm{H}})$  
              & $0.17 \pm 0.11$ & $0.09 \pm 0.03$ & 
			  $0.22 \pm 0.09$ (input) & $0.22 \pm 0.09$ (input)  \\
    $\gamma_4^{\rm{H}}$ 
              & $7.6 \pm 6.5$   & $3.5 \pm 2.2$   & $2.3 _{-2.3}^{+5.0}$ & $0.47 \pm 0.36$ \\
\hline
\hline
\end{tabular}
\end{center}
\end{table}

If we assume the $\Delta I=1/2$ rule for the nonmesonic weak decays, then
an extra condition ${R_{n0}}/{R_{p0}} = 2$ is imposed and it reduces the number of the unknowns.
Therefore, we can determine $R_{NJ}$ without using $\Gamma_{NM}({^4_\Lambda}{\rm{H}})$
as an input.
The first two columns of Table~\ref{ta2} show the resulting $R_{NJ}$.
The predicted value of $\Gamma_{NM}({^4_\Lambda}{\rm{H}})$ and the unknown $n/p$
ratio of ${^4_\Lambda}{\rm{H}}$ are also given.
One sees that by imposing the $\Delta I=1/2$ constraint, the ratio of $R_{p0}$ and $R_{p1}$
changes drastically, while the sum $(R_{p0}+3 R_{p1})$ remains constant.
In fact, the order of $R_{p0}$ and $R_{p1}$ is reversed for set II.
One also sees that $\Gamma_{NM}({^4_\Lambda}{\rm{H}})$ for set II  is much smaller 
than the value given in Table II.  

It is easy to prove the following two relations under the $\Delta I=1/2$ constraint:
\begin{eqnarray}
  \frac{R_{n1}}{R_{n0}} &=& \frac{1}{3} \left( \gamma_4^{\rm{H}} - 1 \right),
 \nonumber\\
\frac{R_{p1}}{R_{p0}} &=& \frac{1}{3} \left( \frac{4}{\gamma_4^{\rm{He}} } - 1 \right) .
\label{eq:DIR}
\end{eqnarray}
The first equation gives a new constraint that 
$\gamma_4^{\rm H}$ must be larger than 1 if the $\Delta I=1/2$ rule is
 satisfied.
The second equation indicates that $R_{p1}$ must be larger than $R_{p0}$
 because the $n/p$ ratio of ${^4_\Lambda}{\rm{He}}$ is smaller than 1.
 These may be useful conditions to test whether the $\Delta I=1/2$ rule is
 satisfied or not.
One can easily confirm that these relations are satisfied for our solutions with
the $\Delta I=1/2$ condition.

The conclusion of the phenomenological analyses of the decay rates and $n/p$ ratios
of the $s$-shell hypernuclei is that the current experimental knowledge already suggests
that the $\Delta I=1/2$ rule is not satisfied in the NMWD, although the precise
measurements of the ${^4_\Lambda}{\rm{H}}$ decays are critical to finalize the
conclusion.

%
\section{Roles of the $\sigma$ meson exchange}
The phenomenological analyses in the last section have revealed us
that the new data for $^5_{\Lambda}{\rm{He}}$ reduce ambiguities in
determining the partial decay rates, particularly for $R_{n1}$.
In this section, we introduce a new element, i.e.,  one-sigma exchange (OSE),
in the microscopic model for the $\Lambda N\to NN$.
A possible importance of OSE has been suggested by approaches
in effective field theory for weak baryonic interaction.
There a short-range weak transition with no charge- or spin-dependence
seems to have significant role to reproduce the decay rates and 
the proton asymmetry of NMWD.

We here propose new microscopic models which incorporates OSE:
(1) The meson exchange (ME) model, which contains OPE+OKE+OSE, and 
(2) the extended direct quark (DQ+) model, which consists of DQ+OPE+OKE+OSE.
ME preserves the $\Delta I =1/2$ rule, while DQ+ predicts significant violation of 
the $\Delta I =1/2$ rule.
The latter is induced by the effective four-quark hamiltonian \cite{ITO:NPA}, and 
is a distinct feature of the direct quark interaction.
We will see that both ME and DQ+ can reproduce the current experimental data more or less, 
but they predict differences in NMWDs of the $A=4$ hypernuclei,
${^4_\Lambda}{\rm{He}}$ and ${^4_\Lambda}{\rm{H}}$.

Key quantities that show importance of OSE is the partial decay rates 
$f_N$ and $b_N$.  
In particular, $|f_n|^2$ is the only component of the $R_{n1}$ decay rate and 
therefore can be determined from the experimental data rather directly.

We here determine the weak $\Lambda n \sigma$ vertex parameters,
 $A_\sigma$ and $B_\sigma$.
They are fixed in the following two steps.
(1) We determine the parameter $B_\sigma$ so as to reproduce the $f_n$ and $b_n$
 decay rates.
(2) Then $A_\sigma$ is determined so that the total decay rate
 of ${^5_\Lambda}$He agrees with the recent experimental data \cite{i4}.

\begin{figure}[t]
\centerline{ \epsfxsize=74mm \epsfbox{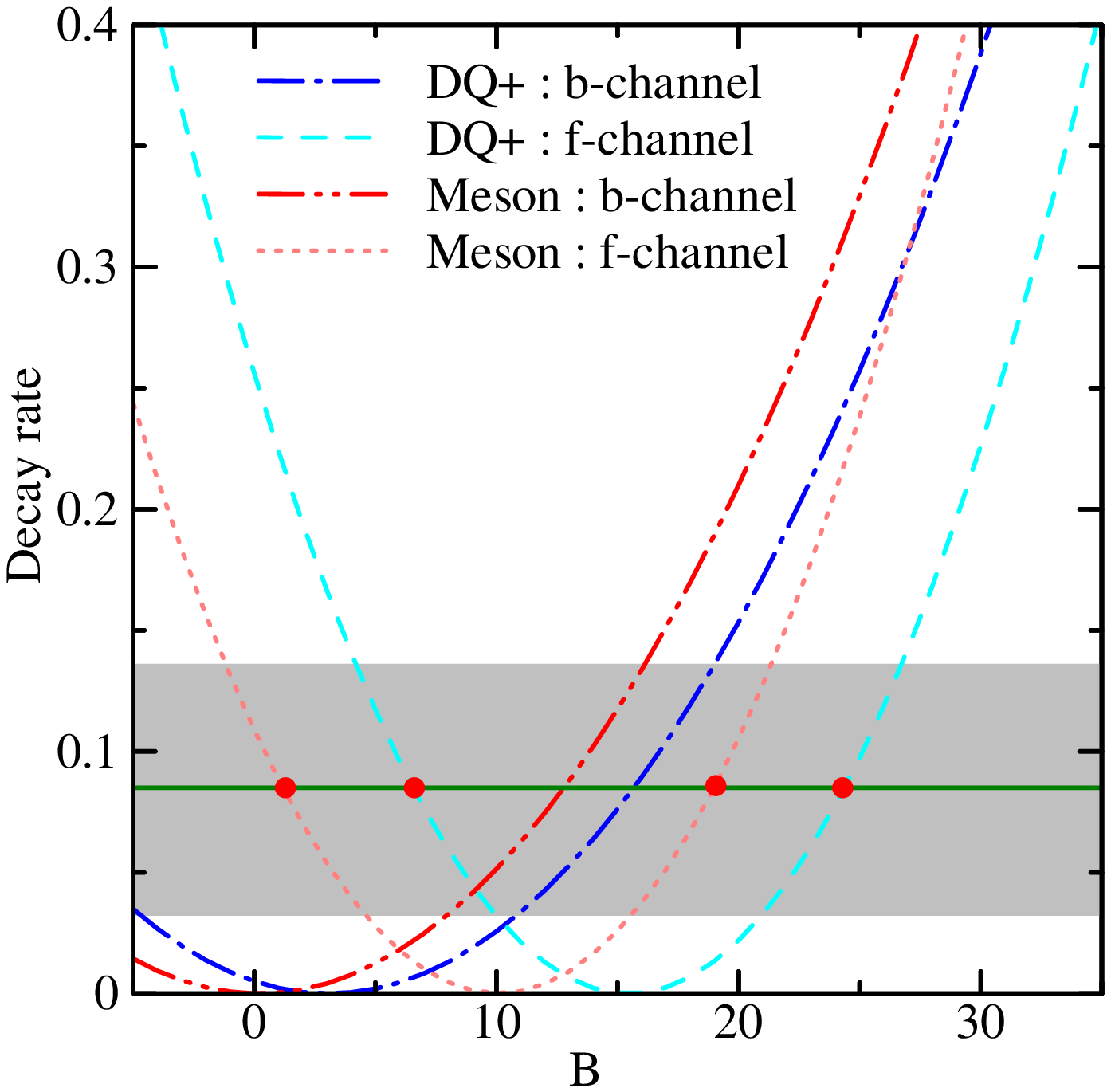} 
             \epsfxsize=67mm \epsfbox{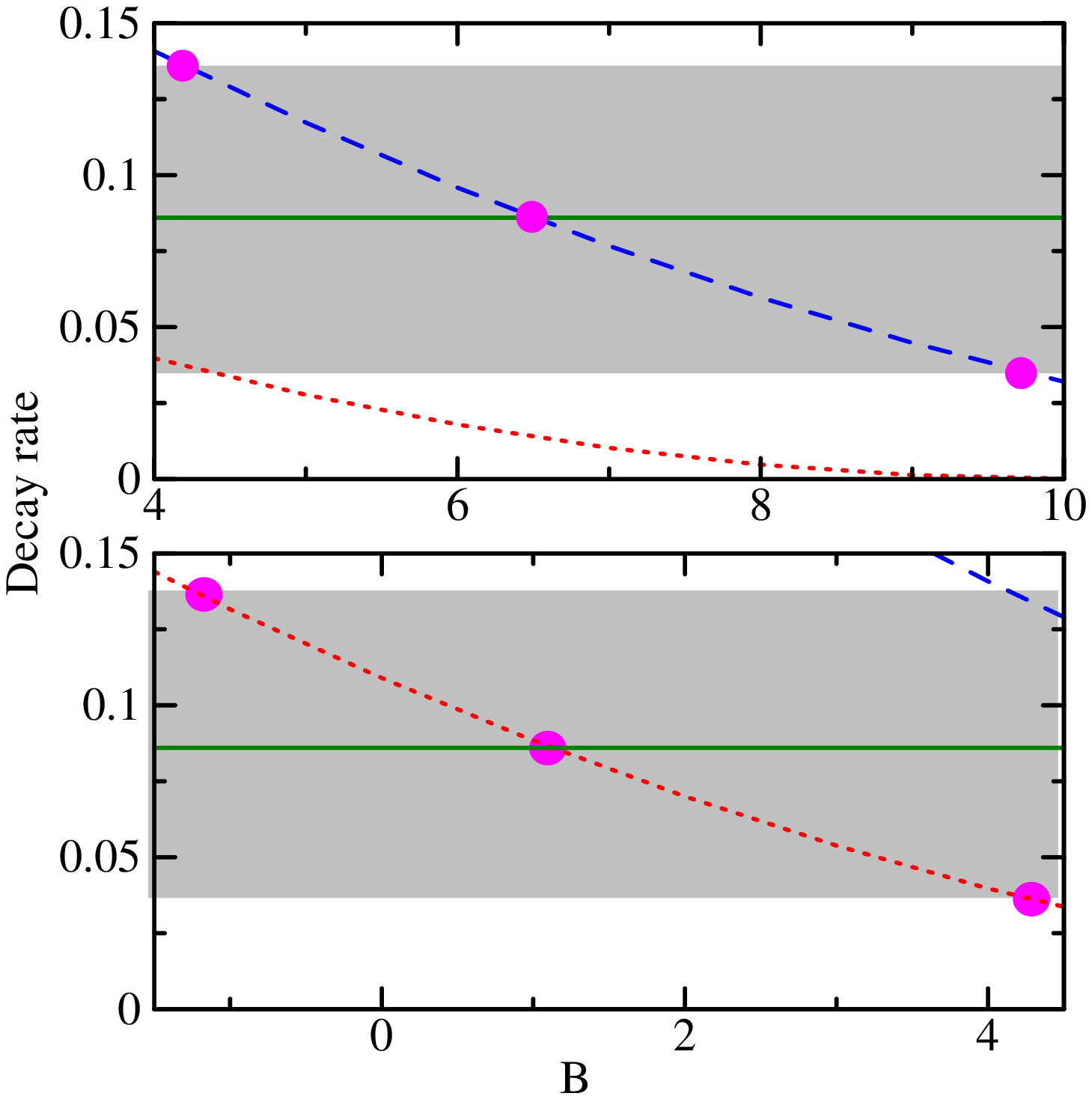}}
\caption{The $B_\sigma$ dependences of the partial decay rates in $b_n$- and $f_n$-channels
          given in units of $\Gamma_\Lambda$.
         The shaded region corresponds to $R_{n1}$ evaluated in
         section 3. The right panels are the enlarged plots around the crossings
         for the $f_n$-channel.}
\label{FIG:bf-ch}
\end{figure}

Fig.~\ref{FIG:bf-ch} shows the $B_\sigma$ dependences of the $b_n$ and $f_n$ decay rates
 for both the ME and DQ+ cases.
The $\Gamma(f_n)$, which is the decay rate of $f_n$-channel,
is quadratic in $B_\sigma$, so that we have two candidates of $B_\sigma$.
This is the channel that has contributions from OPE, OKE and 
DQ added up all coherently and thus plays the central role in solving 
the $n/p$ ratio problem.
Our previous analysis employing the OPE+OKE+DQ model \cite{SIO01:NPA,SIO02:NPA}
was shown to give too much enhancement of $\Gamma(f_n)$ so that 
both the total decay rate and the $n/p$ ratio of ${^5_\Lambda}{\rm{He}}$ are overestimated.
The same enhancement is seen in Fig.~\ref{FIG:bf-ch} at $B_\sigma=0$.
Thus the main role of the parity-violating part of OSE is that it reduces 
$\Gamma(f_n)$ so as to fit $R_{n1}$.

Among the two possibility for $B_\sigma$, the larger one is not appropriate.
This can be seen from the behavior of the other PV decay rate $\Gamma(b_n)$ in Fig.~\ref{FIG:bf-ch}.
If we take the larger $B_\sigma$ (i.e., $\sim 20$), $\Gamma(b_n)$ becomes too large, 
$\sim 0.3 \Gamma_\Lambda$, to accommodate with the observed total decay rates.
Thus we find the ranges for possible $B_{\sigma}$ in the ME and DQ+ as
\begin{eqnarray}
 B_\sigma = \left\{ \begin{array}{rcrcl}
  -1.2 & \sim & 4.4 & : & \hbox{ for ME} \\
   4.2 & \sim & 9.8 & : & \hbox{ for DQ+}
 \end{array} \right.
\end{eqnarray}
In fact, the central value of $R_{n1}$ is reproduced by 
$B_\sigma = 1.2$ for ME and $B_\sigma = 6.6$ for DQ+.

\begin{figure}[t]
\centerline{ \epsfxsize=70mm \epsfbox{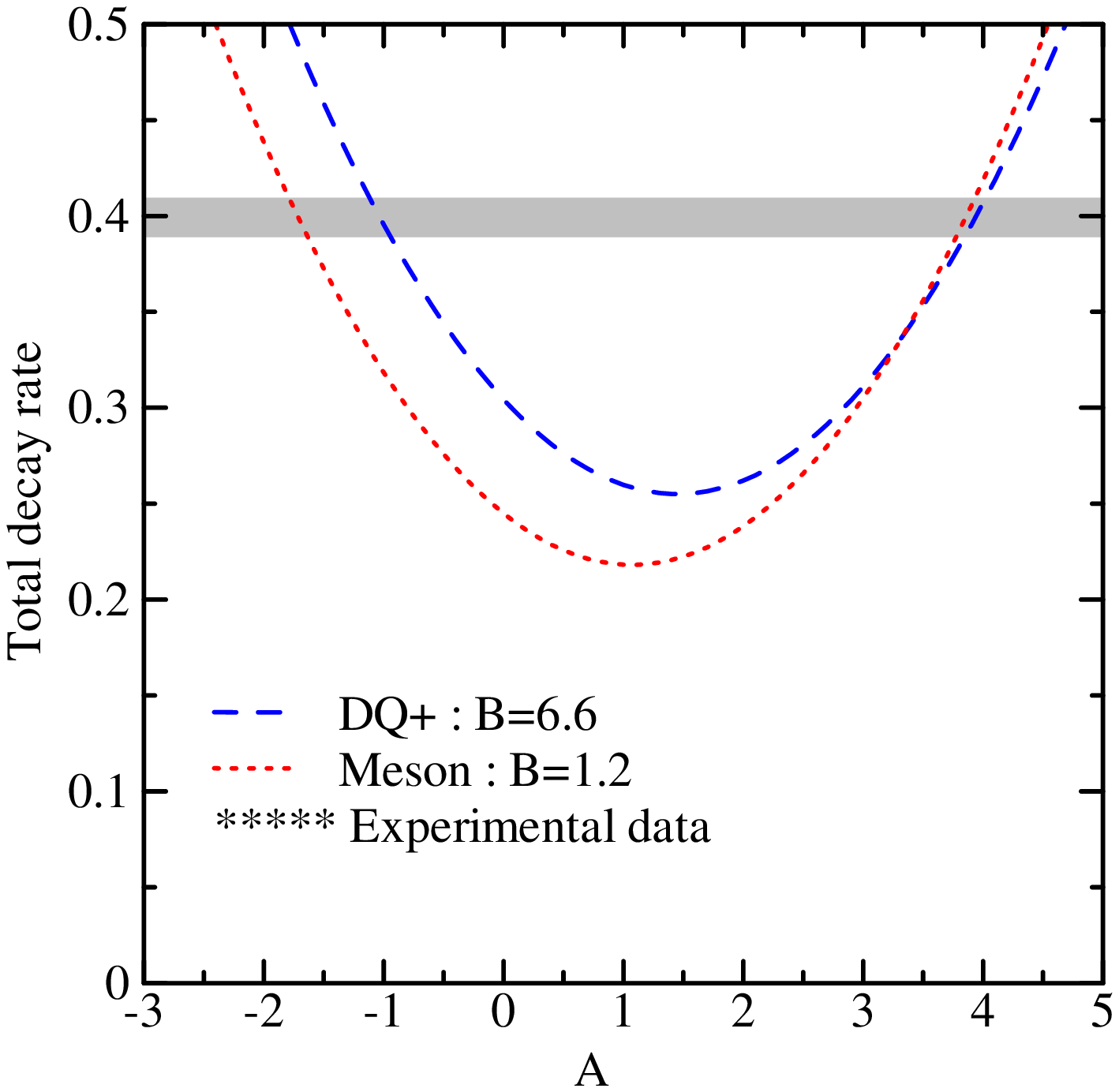} 
             \epsfxsize=71mm \epsfbox{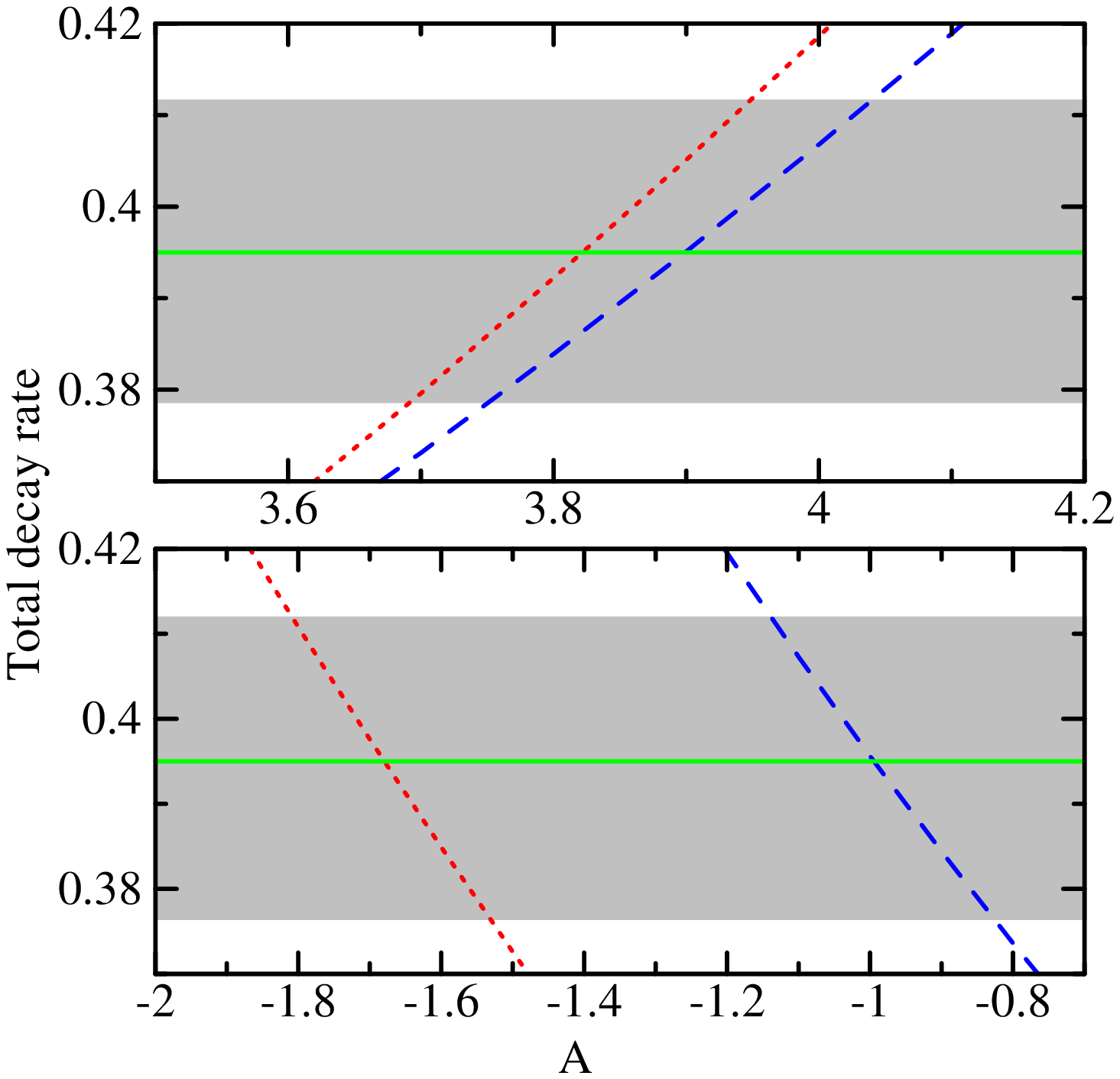}
           }
\caption{The $A_\sigma$ dependence of the total NM decay rate of ${^5_\Lambda}{\rm{He}}$
         in ME and DQ+ models given in units of $\Gamma_\Lambda$.
         The shaded region stands for the experimental value \cite{i4} with the error bar.
         Right two panels are enlarged figure around the intersection 
           with the experimental value.}
\label{FIG:total}
\end{figure}

Fig.~\ref{FIG:total} shows the $A_\sigma$ dependence of total NM decay rate of
 ${^5_\Lambda}{\rm{He}}$ at the central value of $B_\sigma$.
 Because $a_N$ and $c_N$  depend on $A_\sigma$ linearly,
 the total decay rate is a quadratic function of $A_\sigma$.
Thus we again have two candidates of $A_{\sigma}$ given by
\begin{eqnarray}
 A_\sigma = \left\{ \begin{array}{rclcl}
 3.8 & {\rm{and}} & -1.7 & : & \hbox{ for ME} \\
 3.9 & {\rm{and}} & -1.0 & : & \hbox{ for DQ+} 
 \end{array} \right.
\end{eqnarray}

\begin{figure}[t]
\centerline{
 \epsfxsize=70mm \epsfbox{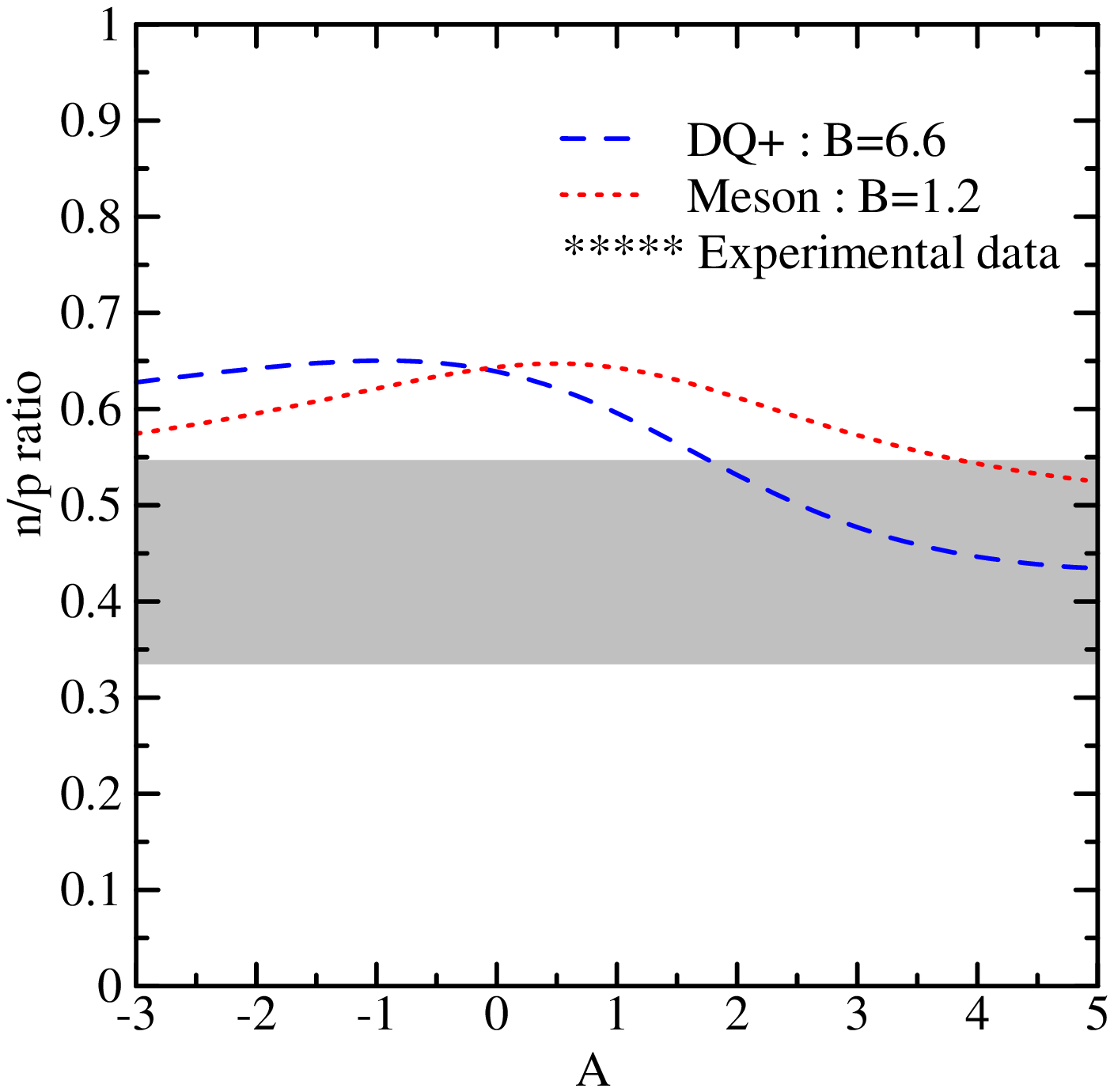}
 \epsfxsize=72mm \epsfbox{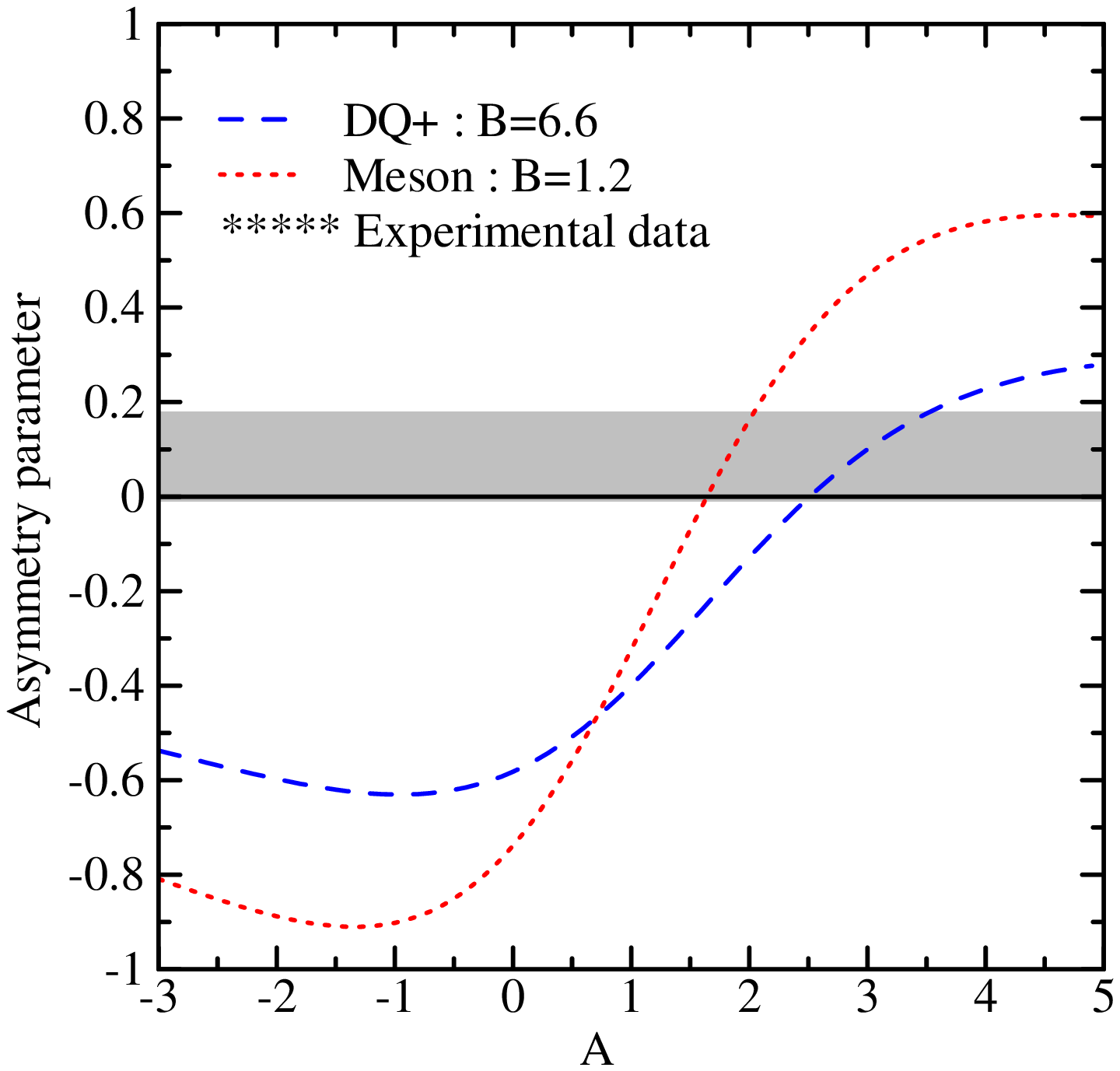}
           }
\caption{The $A_\sigma$ dependences of the $n/p$ ratio and the asymmetry
parameter in ME and DQ+.
         The shaded region stands for the experimental values \cite{i4} with the error bar.}
\label{FIG:np-asym}
\end{figure}

Next in Fig.~\ref{FIG:np-asym} we show the $A_\sigma$ dependence of the $n/p$ ratio 
and the asymmetry parameter, $\alpha$,
 of the NMWD of ${^5_\Lambda}{\rm{He}}$.
One sees in the left panel that the $n/p$ ratio hits the peak at $A_\sigma =1$ for ME, while
the same value gives the minimum of the total NM decay rate of ${^5_\Lambda}{\rm{He}}$.
In contrast, for the DQ+ case, the maximum is given at $A_\sigma =-1$ and the minimum 
appears at $A_\sigma =4$.
For both ME and DQ+,
positive $A_\sigma$ around $4.0$ gives a lower $n/p$ ratio
that is consistent with the experimental data.

The right panel of Fig.~\ref{FIG:np-asym} shows 
the asymmetry parameter, $\alpha$.
We find that it is sensitive to the choice of $A_\sigma$.
For both ME and DQ+, the $\alpha$ becomes large and negative around $A_\sigma =-1$, 
while it is positive at around $A_\sigma = 4$.
The value rises rather rapidly at around $A_\sigma =1$.
The current experimental data for $\alpha$ is small but positive,
and therefore favors positive $A_\sigma$.
%

\begin{figure}[t]
\centerline{
 \epsfxsize=72mm \epsfbox{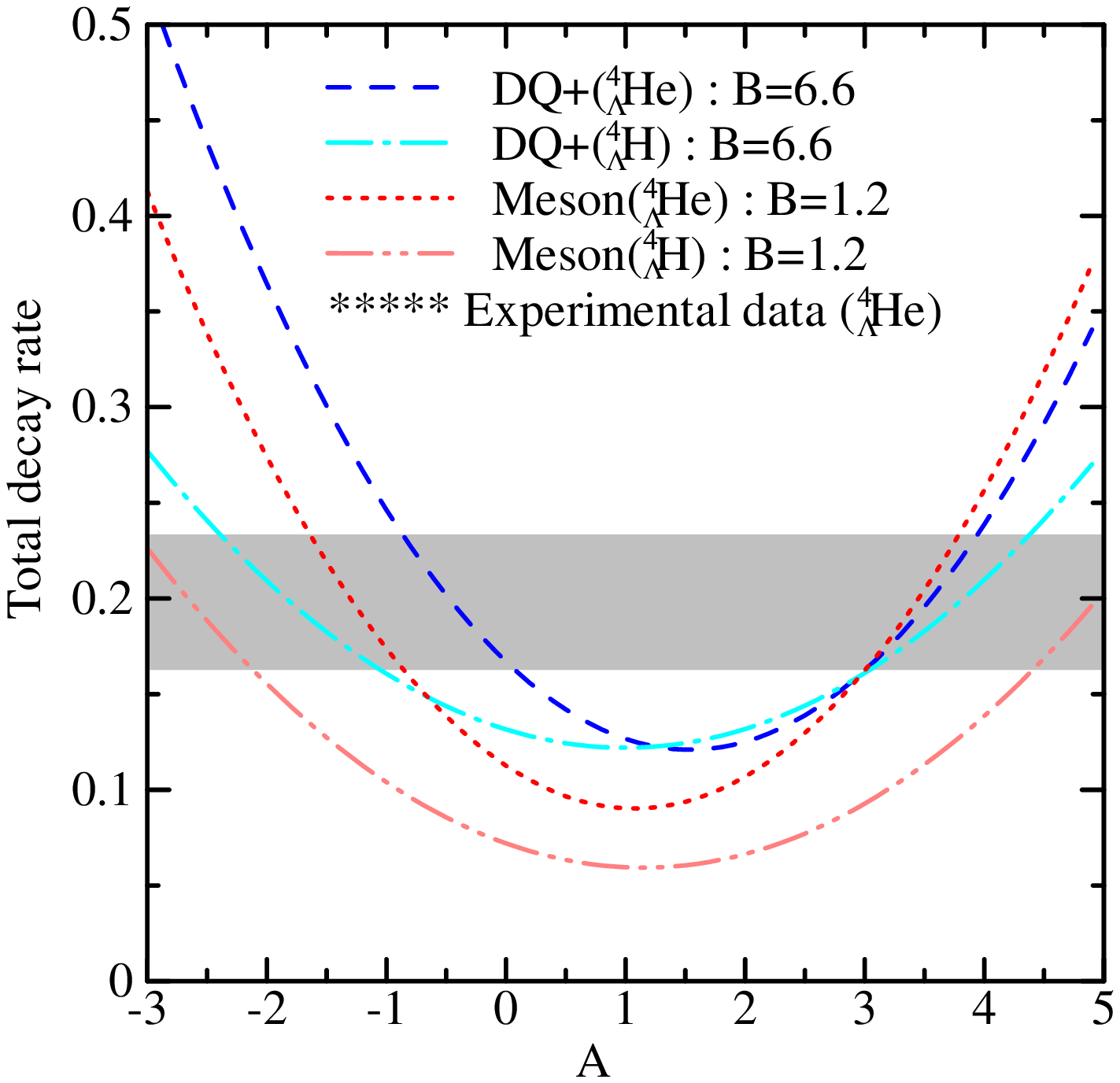}
 \epsfxsize=70mm \epsfbox{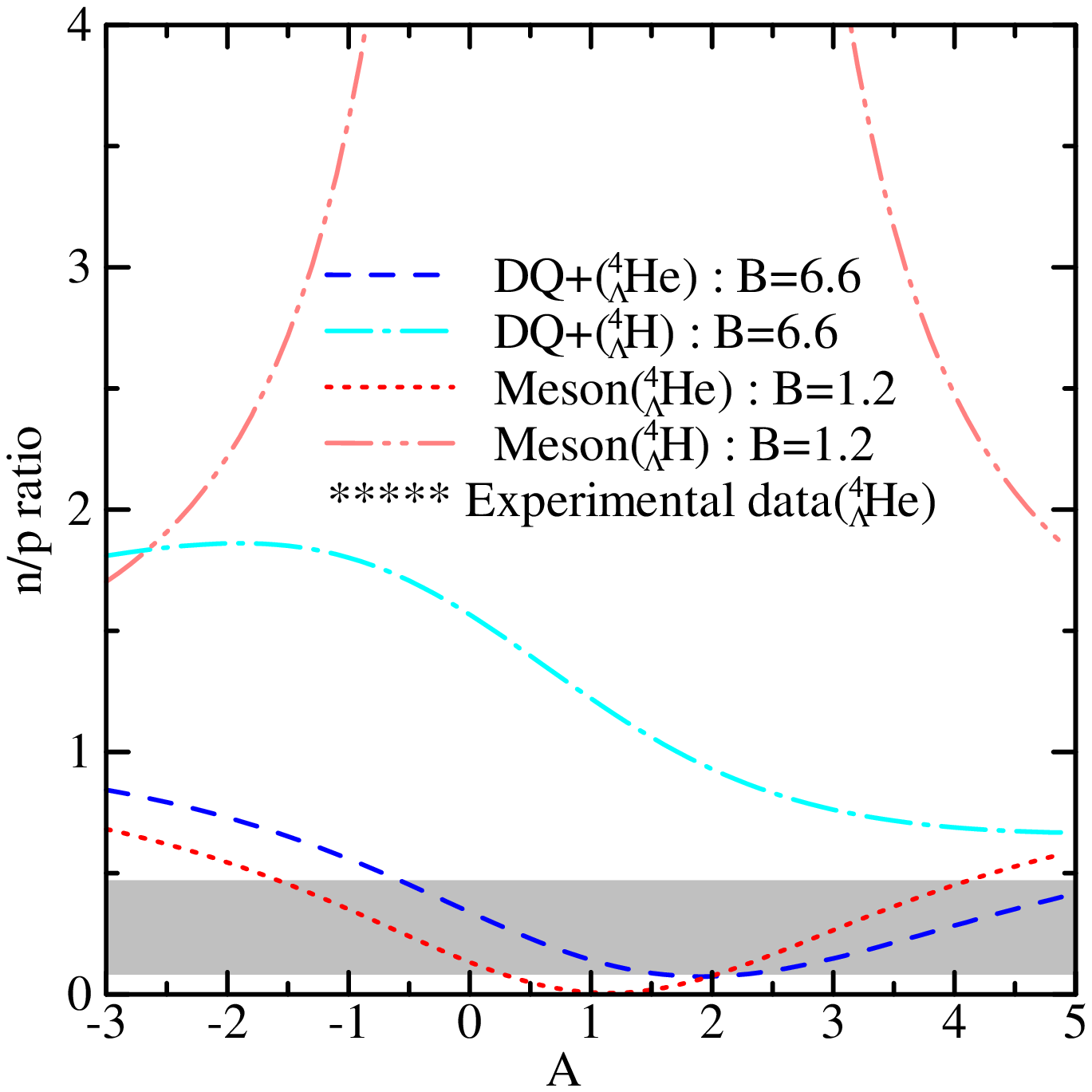}
           }
\caption{The total decay rates and $n/p$ ratios of four-body hypernuclei
          are calculated by meson and DQ+ models in unit of $\Gamma_\Lambda$.
         The shaded region stands for the experimental value of ${^4_\Lambda}{\rm{He}}$
         \cite{Szy:PRC} with error bar.}
\label{FIG:four-body}
\end{figure}

The observables of $A=4$ hypernuclei are also important to understand
 the $\Lambda N \to NN$ weak interactions, especially in the $J=0$ transition channels.
Fig.~\ref{FIG:four-body} shows the $A_\sigma$ dependences of
 the total NM decay rates and $n/p$ ratios of  both ${^4_\Lambda}{\rm{He}}$
 and ${^4_\Lambda}{\rm{H}}$.
The experimental data of ${^4_\Lambda}{\rm{He}}$ are also shown in the figure.
One sees that the total NM decay rate and the $n/p$ ratio 
 of ${^4_\Lambda}{\rm{He}}$ are reproduced
 within the experimental error bar at $A_\sigma =4$ for both ME and DQ+.


The $n/p$ ratio of ${^4_\Lambda}{\rm{H}}$ is interesting
 because it shows clear difference between ME and DQ+.
For ME, 
 the $n/p$ ratio has a huge peak at $A_\sigma = 1$, where the $a$-channel 
 decay rate is extremely small.  It is also seen that this ratio never falls lower than $1$
which is consistent with the $\Delta I=1/2$ condition given in the previous section.
In contrast,
 the $n/p$ ratio calculated by DQ+ model can be lower than $1$, and
 it becomes $0.7$ around $A_\sigma = 4$.
Therefore the $n/p$ ratio of ${^4_\Lambda}{\rm{H}}$ is  
 a key quantity to determine the property of the $\Lambda N \to NN$ weak interaction.

\begin{table}
\caption{The nonmesonic decay rates, $\Gamma_{NM}$, 
the n/p ratios, $\gamma$, and the proton decay asymmetry parameter, $\alpha$, 
predicted in the ME model.  The decay rates are given
in units of $\Gamma_\Lambda$.}
\begin{center}
\begin{tabular}{|cc|cc|cc|cc|c|}
\hline \hline
 & $A_\sigma$ &  3.0 & -0.8 & 3.8 & -1.7 & 4.5 & -2.3 &     \\
 & $B_\sigma$ & -1.2 &      & 1.2 &      & 4.4 &      & EXP \\
\hline
 & $\Gamma_{NM}$ & 0.405 &  0.400 & 0.392 &  0.398 & 0.407 &  0.398 & 0.395 $\pm$ 0.016 \\
${^5_\Lambda}{\rm{He}}$
 & $\gamma_5$    & 0.675 &  0.721 & 0.548 &  0.603 & 0.472 &  0.553 & 0.44 $\pm$ 0.11 \\
 & $\alpha$      & 0.536 & -0.857 & 0.571 & -0.903 & 0.364 & -0.684 & 0.07 $\pm$ 0.08 \\
\hline
${^4_\Lambda}{\rm{He}}$
 & $\Gamma_{NM}$ & 0.199 &  0.195 & 0.235 &  0.240 & 0.298 &  0.291 & 0.20 $\pm$ 0.03 \\
 & $\gamma_4^{\rm He}$ 
                 & 0.219 &  0.249 & 0.417 &  0.492 & 0.692 &  0.781 & 0.25 $\pm$ 0.16 \\
\hline
${^4_\Lambda}{\rm{H}}$
 & $\Gamma_{NM}$ & 0.132 &  0.135 & 0.128 &  0.138 & 0.145 &  0.151 & 0.22 $\pm$ 0.09 \\
 & $\gamma_4^{\rm H}$ 
                 & 6.400 &  5.946 & 2.705 &  2.488 & 1.379 &  1.362 & --- \\
\hline \hline
\end{tabular}
\end{center}
\label{TAB:He5-PKS}
\end{table}

The results of the parameter searches in the ME model are summarized in Table~\ref{TAB:He5-PKS}.
We take three values for the parameter $B_\sigma$, corresponding to 
 the upper, central, and lower values for $R_{n1}$, respectively.  
Then two solutions for $A_\sigma$ are given for each $B_\sigma$ 
 because the total decay rate of ${^5_\Lambda}$He is a quadratic function of $A_\sigma$.
Table~\ref{TAB:He5-PKS} shows that
 the main difference between the positive and negative $A_\sigma$
 appears only in the asymmetry parameter, $\alpha$, and its experimental value
 prefers the positive $A_\sigma$.
We find that 
 the $\gamma_5$, $n/p$ ratio of ${^5_\Lambda}{\rm{He}}$,
 prefers the smaller $R_{n1}$ (the larger $B_\sigma$), while for the $\gamma_4^{\rm He}$
 the larger $R_{n1}$ (the smaller $B_\sigma$) is favorable.

The result for $A_\sigma = 3.8$ and $B_\sigma = 1.2$ give reasonable account of
 most of the observables except the asymmetry parameter $\alpha$.
It is found that
 the total decay rate of ${^4_\Lambda}$H is about a half of ${^4_\Lambda}$He
 and the $n/p$ ratio is about $2.7$.
One can easily check that these values satisfy the conditions 
 for $\Delta I=1/2$ given in the previous section.

\begin{table}
\caption{The nonmesonic decay rates, $\Gamma_{NM}$, 
the n/p ratios, $\gamma$, and the proton decay asymmetry parameter, $\alpha$, 
predicted in the DQ+ model.  The decay rates are given
in units of $\Gamma_\Lambda$.}
\begin{center}
\begin{tabular}{|cc|cc|cc|cc|c|}
\hline \hline
 & $A_\sigma$ & 3.1 & -0.2 & 3.9 & -1.0 & 4.4 & -1.5 &     \\
 & $B_\sigma$ & 4.2 &      & 6.6 &      & 9.8 &      & EXP \\
\hline
 & $\Gamma_{NM}$ & 0.397 &  0.398 & 0.395 &  0.396 &  0.401 &  0.401 & 0.395 $\pm$ 0.016 \\
${^5_\Lambda}{\rm{He}}$
 & $\gamma_5$    & 0.593 &  0.750 & 0.449 &  0.650 &  0.367 &  0.585 & 0.44 $\pm$ 0.11 \\
 & $\alpha$      & 0.248 & -0.640 & 0.219 & -0.630 & -0.005 & -0.400 & 0.07 $\pm$ 0.08 \\
\hline
${^4_\Lambda}{\rm{He}}$
 & $\Gamma_{NM}$ & 0.184 &  0.196 & 0.229 &  0.246 &  0.288 &  0.308 & 0.20 $\pm$ 0.03 \\
 & $\gamma_4^{\rm He}$      
                 & 0.091 &  0.274 & 0.269 &  0.559 &  0.498 &  0.870 & 0.25 $\pm$ 0.16 \\
\hline
${^4_\Lambda}{\rm{H}}$
 & $\Gamma_{NM}$ & 0.179 &  0.150 & 0.204 &  0.161 &  0.244 &  0.192 & 0.22 $\pm$ 0.09 \\
 & $\gamma_4^{\rm H}$      
                 & 1.396 &  3.649 & 0.693 &  1.802 &  0.411 &  0.979 & --- \\
\hline \hline
\end{tabular}
\end{center}
\label{TAB:He5-PKDQS}
\end{table}

Table~\ref{TAB:He5-PKDQS} shows the results for the DQ+ model.
Again two solutions for $A_{\sigma}$ can reproduce the total decay rate of $^5_\Lambda$He, 
 but the positive $A_\sigma$ explains all the available experimental data 
 for both $A=4$ and $5$ hypernuclei fairly well.
The negative $A_\sigma$ tends to overestimate the $n/p$ ratio of all hypernuclear systems,
and therefore this choice is ruled out.

The calculation with $A_\sigma = 3.9$ and $B_\sigma = 6.6$ gives the best agreement
 with all experimental data among other parameter sets.
In particular, we note that the proton asymmetry parameter
 for ${^5_\Lambda}$He is predicted to be positive and small in this model.
It is brought mainly by OSE, which gives a major
 contribution to the $J=0$ amplitudes.
The $\sigma$ exchange potential changes the sign of the $a$- and $c$-amplitudes
 from those without OSE and, thus, it leads to the drastic change
 of the proton asymmetry parameter, $\alpha$.

The DQ+ model has a prominent feature which is the strong violation
 of the $\Delta I=1/2$ rule.
It can be easily demonstrated from the value of $\kappa$,
\begin{eqnarray}
 \kappa = \frac{\Gamma_{n}({^4_\Lambda}{\rm{He}})}{\Gamma_{p}({^4_\Lambda}{\rm{H}})}
        = 0.42.
\end{eqnarray}
Because $\Delta I=1/2$ will lead to $\kappa=2$, this result indicates 
 a large violation of the $\Delta I=1/2$ rule due to the DQ contribution.
It is also seen from 
 the total NM decay rates of ${^4_\Lambda}$He and ${^4_\Lambda}$H, which are almost equal,
 and the $n/p$ ratios for $A=4$, which are less than $1$.
As is shown in the previous section,
 these properties also indicate a large $\Delta I=3/2$ contribution.
 
\begin{figure}[t]
\centerline{
 \epsfxsize=70mm \epsfbox{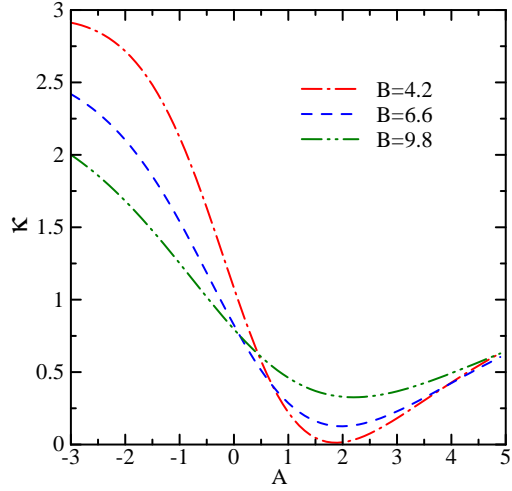}
           }
\caption{The $A_\sigma$ and $B_\sigma$ dependences of $\kappa$
         in the DQ+ model.
         If the $\Delta I=1/2$ rule is preserved,
          the $\kappa$ is equal to $2$.}
\label{FIG:kappa}
\end{figure}

It is interesting to see the $A_\sigma$ and $B_\sigma$ dependences of $\kappa$,
 which is shown in Fig~\ref{FIG:kappa}.
For $B_\sigma =4.2$ case that corresponds to the upper limit of $R_{n1}$,
$\kappa$ hardly change with the $A_\sigma$.
All the curves cross the $\kappa =2$ line in the region of negative $A_\sigma$.
Therefore
 it must be noted that there is possibility to observe $\kappa = 2$ accidentally,
 even if the $\Delta I=1/2$ rule is largely broken in a microscopic interaction.
However,
 the model which we recommend chooses the $A_\sigma$ around $4.0$,
 so that we can observe $\kappa$ around $0.5$ and the violation of $\Delta I =1/2$ rule
 can be seen.

In total, we conclude that the overall agreement of the theoretical predictions
 with experimental data in the DQ+ model is much better than the ME model.
This suggests strongly that the violation of $\Delta I=1/2$ rule is also favored by
 the current data set, although the definite conclusion will be given only after a future
 precise measurement of the NM decay of ${^4_\Lambda}{\rm{H}}$.

%
\section{Conclusions}

A microscopic picture of the $\Lambda N \to NN$ weak interaction 
has been established including exchange of a scalar-isoscalar meson, $\sigma$, i.e.,
one-scalar-exchange (OSE) interaction.
Our full model, called DQ+ model, consists of the short-range DQ interaction as well as the long-range
$\pi+K+\sigma$ exchange interactions.
We have found that the new data for ${^5_\Lambda}$He are very powerful in
determining the weak $\sigma N \Lambda$ coupling constants.
They reduce ambiguity of the $R_{n1}$ decay rate, which in turn determines
the PV $\sigma N \Lambda$ coupling, $B_\sigma$, rather precisely.
The PC part, $A_\sigma$, can then be determined by the total NM decay rate
of  ${^5_\Lambda}$He.

The established DQ+ model has been shown to reproduce all the current experimental data 
of four- and five-body hypernuclei fairly well.
In particular, the asymmetry parameter of the proton emitted from polarized ${^5_\Lambda}$He
is now consistent with recent experimental data.

A parallel analysis by the meson exchange (ME) model without the DQ part of the interaction 
 is also found to explain most of the experimental data except for the asymmetry parameter,
 although the fit seems better in DQ+.
The main difference between ME and DQ+ is its isospin property.
The ME interactions preserve the $\Delta I=1/2$ rule
and predicts a small decay rate and a large $n/p$ ratio in the NM decay of ${^4_\Lambda}{\rm{H}}$.
The DQ+ model, on the other hand, shows a larger decay rate, comparable to that of the NM
decay of ${^4_\Lambda}{\rm{He}}$, and a smaller $n/p$ ratio in the ${^4_\Lambda}{\rm{H}}$ decay. 
Our analysis shows that the DQ+ model introduces a significant $\Delta I=3/2$ contribution
brought by the effective four-quark hamiltonian, and thus predicts violation of the $\Delta I=1/2$ rule.
Crude estimates from the present knowledge on the ${^4_\Lambda}{\rm{H}}$ decay show 
a large NM decay rate and thus supports the violation of the $\Delta I=1/2$ rule.

We again stress that a direct measurement of the ${^4_\Lambda}{\rm{H}}$ decay is 
indispensable to establish the violation of the $\Delta I=1/2$ rule and hope that such 
experiment is realized in the near future.

\section*{Acknowledgment}
\noindent
One of authors, K. S., acknowledges JSPS Research Fellowship for financial support.
This work is supported in part by the Grant for Scientific Research (B)No.15340072 and (C)No.16540236
from the Ministry of Education, Culture, Sports, Science and Technology, Japan.

\end{document}